\documentstyle[12pt,dina4]{article}
\def\title #1 {
   \headsep=1.0in
   \baselineskip=30pt
			\begin{center}
   {\titlebold #1}
   \end{center}
			\vskip .75in }
\def\author #1 {
   \baselineskip=30pt
   \begin{center}
   {\timeslarge #1}
   \end{center}
			\vskip .25in }
\def\address #1 {
   \baselineskip=24pt
   \begin{center}
   {\timesitalic #1}
   \end{center}
   \vskip 1.0in }

\def\disp {\displaystyle}

\def\conj #1 {\overline #1}

\def\be {\begin{equation}}
\def\ee {\end{equation}}

\def\ba {\begin{array}}
\def\ea {\end{array}}
\def\bea {\begin{eqnarray}}
\def\eea {\end{eqnarray}}

\def\et {$$}

\def\etn {$$}

\def\ett {$$}

\def\ettn{$$}

\def\eqn#1 {(\ref{#1}) }

\newdimen\twoeqncolwidth
\newdimen\twoeqncolwidtha
\newdimen\twoeqncolwidthb
\newdimen\twoeqncolsep
\newdimen\twoeqnlinset
\def\twoeqn#1&#2\et{
   \hbox to\twoeqnlinset{\hfil}
   \hbox to\twoeqncolwidth{$\disp#1$\hfil}
   \hbox to\twoeqncolsep{\hfil}
   \hbox to\twoeqncolwidth{$\disp#2$\hfil}\eqno{\rm (\theequation)}$$}
\def\twoeqnt#1&#2\ett{
   \hbox to\twoeqnlinset{\hfil}
   \hbox to\twoeqncolwidtha{$\disp#1$\hfil}
   \hbox to\twoeqncolsep{\hfil}
   \hbox to\twoeqncolwidthb{$\disp#2$\hfil}\eqno{\rm (\theequation)}$$}
\def\twoeqnn#1&#2\etn{
   \hbox to\twoeqnlinset{\hfil}
   \hbox to\twoeqncolwidth{$\disp#1$\hfil}
   \hbox to\twoeqncolsep{\hfil}
   \hbox to\twoeqncolwidth{$\disp#2$\hfil}\eqno\phantom{\rm
(\theequation)}$$}
\def\twoeqntn#1&#2\ettn{
   \hbox to\twoeqnlinset{\hfil}
   \hbox to\twoeqncolwidtha{$\disp#1$\hfil}
   \hbox to\twoeqncolsep{\hfil}
   \hbox to\twoeqncolwidthb{$\disp#2$\hfil}\eqno\phantom{\rm
(\theequation)}$$}

\def\rawpicture #1 by #2 (#3){
  \vbox to #2{
    \hrule width #1 height 0pt depth 0pt
    \vfill
    }
  }
\def\scaledpicture #1 by #2 (#3 scaled #4){{
  \dimen0=#1 \dimen1=#2
  \divide\dimen0 by 1000 \multiply\dimen0 by #4
  \divide\dimen1 by 1000 \multiply\dimen1 by #4
  \rawpicture \dimen0 by \dimen1 (#3 scaled #4)}
  }

\renewcommand{\title}[1]{\large\bf \mbox{}\\ \mbox{}\\ \mbox{}\\ \mbox{}\\
     #1\bigskip\medskip\\}
\renewcommand{\author}[1]{\large #1\\ \smallskip}
\renewcommand{\address}[1]{{\narrower\normalsize\it #1\\}\bigskip}
\renewenvironment{abstract}{\narrower\small}{\par\normalsize\bigskip}

\catcode`\@=11

\font\twelvemsx=msxm10 scaled \magstep1
\font\tenmsx=msxm10
\font\sevenmsx=msxm7

\font\twelvemsy=msym10 scaled \magstep1
\font\tenmsy=msym10
\font\sevenmsy=msym7

\newfam\msxfam
\newfam\msyfam
\textfont\msxfam=\twelvemsx
\scriptfont\msxfam=\tenmsx
  \scriptscriptfont\msxfam=\sevenmsx
\textfont\msyfam=\twelvemsy
\scriptfont\msyfam=\tenmsy
  \scriptscriptfont\msyfam=\sevenmsy

\def\hexnumber@#1{\ifcase#1 0\or1\or2\or3\or4\or5\or6\or7\or8\or9\or
	A\or B\or C\or D\or E\or F\fi }

\font\twelveeuf=eufm10 scaled \magstep1
\font\teneuf=eufm10
\font\seveneuf=eufm7

\newfam\euffam
\textfont\euffam=\twelveeuf
\scriptfont\euffam=\teneuf
\scriptscriptfont\euffam=\seveneuf
\def\frak{\relaxnext@\ifmmode\let\next\frak@\else
 \def\next{\Err@{Use \string\frak\space only in math mode}}\fi\next}
\def\goth{\relaxnext@\ifmmode\let\next\frak@\else
 \def\next{\Err@{Use \string\goth\space only in math mode}}\fi\next}
\def\frak@#1{{\frak@@{#1}}}
\def\frak@@#1{\noaccents@\fam\euffam#1}

\edef\msx@{\hexnumber@\msxfam}
\edef\msy@{\hexnumber@\msyfam}

\mathchardef\boxdot="2\msx@00
\mathchardef\boxplus="2\msx@01
\mathchardef\boxtimes="2\msx@02
\mathchardef\square="0\msx@03
\mathchardef\blacksquare="0\msx@04
\mathchardef\centerdot="2\msx@05
\mathchardef\lozenge="0\msx@06
\mathchardef\blacklozenge="0\msx@07
\mathchardef\circlearrowright="3\msx@08
\mathchardef\circlearrowleft="3\msx@09
\mathchardef\rightleftharpoons="3\msx@0A
\mathchardef\leftrightharpoons="3\msx@0B
\mathchardef\boxminus="2\msx@0C
\mathchardef\Vdash="3\msx@0D
\mathchardef\Vvdash="3\msx@0E
\mathchardef\vDash="3\msx@0F
\mathchardef\twoheadrightarrow="3\msx@10
\mathchardef\twoheadleftarrow="3\msx@11
\mathchardef\leftleftarrows="3\msx@12
\mathchardef\rightrightarrows="3\msx@13
\mathchardef\upuparrows="3\msx@14
\mathchardef\downdownarrows="3\msx@15
\mathchardef\upharpoonright="3\msx@16

\mathchardef\downharpoonright="3\msx@17
\mathchardef\upharpoonleft="3\msx@18
\mathchardef\downharpoonleft="3\msx@19
\mathchardef\rightarrowtail="3\msx@1A
\mathchardef\leftarrowtail="3\msx@1B
\mathchardef\leftrightarrows="3\msx@1C
\mathchardef\rightleftarrows="3\msx@1D
\mathchardef\Lsh="3\msx@1E
\mathchardef\Rsh="3\msx@1F
\mathchardef\rightsquigarrow="3\msx@20
\mathchardef\leftrightsquigarrow="3\msx@21
\mathchardef\looparrowleft="3\msx@22
\mathchardef\looparrowright="3\msx@23
\mathchardef\circeq="3\msx@24
\mathchardef\succsim="3\msx@25
\mathchardef\gtrsim="3\msx@26
\mathchardef\gtrapprox="3\msx@27
\mathchardef\multimap="3\msx@28
\mathchardef\therefore="3\msx@29
\mathchardef\because="3\msx@2A
\mathchardef\doteqdot="3\msx@2B

\mathchardef\triangleq="3\msx@2C
\mathchardef\precsim="3\msx@2D
\mathchardef\lesssim="3\msx@2E
\mathchardef\lessapprox="3\msx@2F
\mathchardef\eqslantless="3\msx@30
\mathchardef\eqslantgtr="3\msx@31
\mathchardef\curlyeqprec="3\msx@32
\mathchardef\curlyeqsucc="3\msx@33
\mathchardef\preccurlyeq="3\msx@34
\mathchardef\leqq="3\msx@35
\mathchardef\leqslant="3\msx@36
\mathchardef\lessgtr="3\msx@37
\mathchardef\backprime="0\msx@38
\mathchardef\risingdotseq="3\msx@3A
\mathchardef\fallingdotseq="3\msx@3B
\mathchardef\succcurlyeq="3\msx@3C
\mathchardef\geqq="3\msx@3D
\mathchardef\geqslant="3\msx@3E
\mathchardef\gtrless="3\msx@3F
\mathchardef\sqsubset="3\msx@40
\mathchardef\sqsupset="3\msx@41
\mathchardef\vartriangleright="3\msx@42
\mathchardef\vartriangleleft="3\msx@43
\mathchardef\trianglerighteq="3\msx@44
\mathchardef\trianglelefteq="3\msx@45
\mathchardef\bigstar="0\msx@46
\mathchardef\between="3\msx@47
\mathchardef\blacktriangledown="0\msx@48
\mathchardef\blacktriangleright="3\msx@49
\mathchardef\blacktriangleleft="3\msx@4A
\mathchardef\vartriangle="0\msx@4D
\mathchardef\blacktriangle="0\msx@4E
\mathchardef\triangledown="0\msx@4F
\mathchardef\eqcirc="3\msx@50
\mathchardef\lesseqgtr="3\msx@51
\mathchardef\gtreqless="3\msx@52
\mathchardef\lesseqqgtr="3\msx@53
\mathchardef\gtreqqless="3\msx@54
\mathchardef\Rrightarrow="3\msx@56
\mathchardef\Lleftarrow="3\msx@57
\mathchardef\veebar="2\msx@59
\mathchardef\barwedge="2\msx@5A
\mathchardef\doublebarwedge="2\msx@5B
\mathchardef\angle="0\msx@5C
\mathchardef\measuredangle="0\msx@5D
\mathchardef\sphericalangle="0\msx@5E
\mathchardef\varpropto="3\msx@5F
\mathchardef\smallsmile="3\msx@60
\mathchardef\smallfrown="3\msx@61
\mathchardef\Subset="3\msx@62
\mathchardef\Supset="3\msx@63
\mathchardef\Cup="2\msx@64

\mathchardef\Cap="2\msx@65

\mathchardef\curlywedge="2\msx@66
\mathchardef\curlyvee="2\msx@67
\mathchardef\leftthreetimes="2\msx@68
\mathchardef\rightthreetimes="2\msx@69
\mathchardef\subseteqq="3\msx@6A
\mathchardef\supseteqq="3\msx@6B
\mathchardef\bumpeq="3\msx@6C
\mathchardef\Bumpeq="3\msx@6D
\mathchardef\lll="3\msx@6E

\mathchardef\ggg="3\msx@6F

\mathchardef\circledS="0\msx@73
\mathchardef\pitchfork="3\msx@74
\mathchardef\dotplus="2\msx@75
\mathchardef\backsim="3\msx@76
\mathchardef\backsimeq="3\msx@77
\mathchardef\complement="0\msx@7B
\mathchardef\intercal="2\msx@7C
\mathchardef\circledcirc="2\msx@7D
\mathchardef\circledast="2\msx@7E
\mathchardef\circleddash="2\msx@7F
\def\ulcorner{\delimiter"4\msx@70\msx@70 }
\def\urcorner{\delimiter"5\msx@71\msx@71 }
\def\llcorner{\delimiter"4\msx@78\msx@78 }
\def\lrcorner{\delimiter"5\msx@79\msx@79 }
\def\yen{\mathhexbox\msx@55 }
\def\checkmark{\mathhexbox\msx@58 }
\def\circledR{\mathhexbox\msx@72 }
\def\maltese{\mathhexbox\msx@7A }
\mathchardef\lvertneqq="3\msy@00
\mathchardef\gvertneqq="3\msy@01
\mathchardef\nleq="3\msy@02
\mathchardef\ngeq="3\msy@03
\mathchardef\nless="3\msy@04
\mathchardef\ngtr="3\msy@05
\mathchardef\nprec="3\msy@06
\mathchardef\nsucc="3\msy@07
\mathchardef\lneqq="3\msy@08
\mathchardef\gneqq="3\msy@09
\mathchardef\nleqslant="3\msy@0A
\mathchardef\ngeqslant="3\msy@0B
\mathchardef\lneq="3\msy@0C
\mathchardef\gneq="3\msy@0D
\mathchardef\npreceq="3\msy@0E
\mathchardef\nsucceq="3\msy@0F
\mathchardef\precnsim="3\msy@10
\mathchardef\succnsim="3\msy@11
\mathchardef\lnsim="3\msy@12
\mathchardef\gnsim="3\msy@13
\mathchardef\nleqq="3\msy@14
\mathchardef\ngeqq="3\msy@15
\mathchardef\precneqq="3\msy@16
\mathchardef\succneqq="3\msy@17
\mathchardef\precnapprox="3\msy@18
\mathchardef\succnapprox="3\msy@19
\mathchardef\lnapprox="3\msy@1A
\mathchardef\gnapprox="3\msy@1B
\mathchardef\nsim="3\msy@1C
\mathchardef\ncong="3\msy@1D

\mathchardef\varsubsetneq="3\msy@20
\mathchardef\varsupsetneq="3\msy@21
\mathchardef\nsubseteqq="3\msy@22
\mathchardef\nsupseteqq="3\msy@23
\mathchardef\subsetneqq="3\msy@24
\mathchardef\supsetneqq="3\msy@25
\mathchardef\varsubsetneqq="3\msy@26
\mathchardef\varsupsetneqq="3\msy@27
\mathchardef\subsetneq="3\msy@28
\mathchardef\supsetneq="3\msy@29
\mathchardef\nsubseteq="3\msy@2A
\mathchardef\nsupseteq="3\msy@2B
\mathchardef\nparallel="3\msy@2C
\mathchardef\nmid="3\msy@2D
\mathchardef\nshortmid="3\msy@2E
\mathchardef\nshortparallel="3\msy@2F
\mathchardef\nvdash="3\msy@30
\mathchardef\nVdash="3\msy@31
\mathchardef\nvDash="3\msy@32
\mathchardef\nVDash="3\msy@33
\mathchardef\ntrianglerighteq="3\msy@34
\mathchardef\ntrianglelefteq="3\msy@35
\mathchardef\ntriangleleft="3\msy@36
\mathchardef\ntriangleright="3\msy@37
\mathchardef\nleftarrow="3\msy@38
\mathchardef\nrightarrow="3\msy@39
\mathchardef\nLeftarrow="3\msy@3A
\mathchardef\nRightarrow="3\msy@3B
\mathchardef\nLeftrightarrow="3\msy@3C
\mathchardef\nleftrightarrow="3\msy@3D
\mathchardef\divideontimes="2\msy@3E
\mathchardef\varnothing="0\msy@3F
\mathchardef\nexists="0\msy@40
\mathchardef\mho="0\msy@66
\mathchardef\eth="0\msy@67
\mathchardef\eqsim="3\msy@68
\mathchardef\beth="0\msy@69
\mathchardef\gimel="0\msy@6A
\mathchardef\daleth="0\msy@6B
\mathchardef\lessdot="3\msy@6C
\mathchardef\gtrdot="3\msy@6D
\mathchardef\ltimes="2\msy@6E
\mathchardef\rtimes="2\msy@6F
\mathchardef\shortmid="3\msy@70
\mathchardef\shortparallel="3\msy@71
\mathchardef\smallsetminus="2\msy@72
\mathchardef\thicksim="3\msy@73
\mathchardef\thickapprox="3\msy@74
\mathchardef\approxeq="3\msy@75
\mathchardef\succapprox="3\msy@76
\mathchardef\precapprox="3\msy@77
\mathchardef\curvearrowleft="3\msy@78
\mathchardef\curvearrowright="3\msy@79
\mathchardef\digamma="0\msy@7A
\mathchardef\varkappa="0\msy@7B
\mathchardef\hslash="0\msy@7D
\mathchardef\hbar="0\msy@7E
\mathchardef\backepsilon="3\msy@7F
\def\Bbb@@#1{\fam\msyfam#1}
\def\Bbb@#1{{\Bbb@@{#1}}}
\def\Bbb{\Bbb@}

\catcode`\@=12

\font\twelvemsx=msxm10 scaled \magstep1
\font\twelvemsy=msym10 scaled \magstep1
\font\twelveeuf=eufm10 scaled \magstep1
\textfont\msxfam=\twelvemsx
\textfont\msyfam=\twelvemsy
\textfont\euffam=\twelveeuf
\def\frak#1{\mbox{\twelveeuf #1}}

\begin{document}
\begin{center}
\titlepage

\title{Exact Groundstates for Antiferromagnetic Spin-One Chains with
Nearest and Next-Nearest Neighbour
Interactions\footnote{\it Work performed within the research program of the
Sonderforschungsbereich 341, K\"oln-\\Aachen-J\"ulich}}

\vskip0.8cm

\author{C.~Lange, A.~Kl\"umper, J.~Zittartz}
\address{Institut f\"ur
Theoretische Physik,
Universit\"at zu K\"oln, Z\"ulpicher Str. 77,\\
D-50937 K\"oln, Germany.
\footnote{Email: kluemper@thp.uni-koeln.de, zitt@thp.uni-koeln.de}}
\end{center}

\vskip0.8cm

\begin{abstract}
\noindent
We have found the exact ground state for a large class of antiferromagnetic
spin-one chains with nearest and next-nearest neighbour interactions.
The ground state is characterized as a matrix product of local site states
and has the properties characteristic of the Haldane scenario.

\end{abstract}

\section{Introduction}
In recent years antiferromagnetic spin-one chains have been subject
to intensive analytical, numerical and experimental investigations.
Of main interest are groundstate phase diagrams and critical properties
with respect to variations of interaction parameters resp. anisotropies
of the systems. The study of anisotropy effects is quite important, as all
experimental realizations indicate that quasi-one-dimensional systems
have restricted symmetries [1-7].\\
\\
In \cite{zit93} we started the investigation of a most general class
of spin-one chains with anisotropic nearest-neighbour
exchange interactions and single ion anisotropy.
In a large parameter subspace of this model the groundstates were found
in the form of ``matrix products". The corresponding phase diagram
consists of several parts which are separated by transition lines
of first and second order. Away from these lines the model has
a unique ground state, an energy gap to the excited states and
exponential decay of ground state correlations. Thus, the so-called
Haldane scenario was verified for the considered model. This scenario,
conjectured in 1983 \cite{hal83a,hal83b}
for certain isotropic models with integral spins,
is quite interesting as it points out a striking difference between
the behaviour of isotropic
integral and half-integral spin chains. The latter are expected to
have no gap and algebraic decay of correlations.\\
\\
The first model for which the Haldane scenario
was proven rigorously is the Valence-Bond-Solid (VBS) model ---
a spin-one chain with special isotropic bilinear and
biquadratic nearest neighbour interactions \cite{affl87,affl88}.
In a further development the unique VBS ground state
was cast into a matrix product of single site states and generalized
to anisotropic interactions \cite{zit91,zit92}.
The concept of ``matrix product ground states" (MPGs)
was then applied to most general spin-one models \cite{zit93}.
The MPGs have non-trivial correlations and thus differ notably from the
ground states of the Majumdar-Ghosh-type models \cite{majum69}.\\
\\
In this paper we investigate the additional effects of interactions between
next-nearest neighbours \cite{cl}. The anisotropies assumed in \cite{zit93} for
the local spin pair
interactions will be retained.\\

\section{Model}
For realistic spin-one chains we assume the following symmetries:\\

a) rotational invariance in the $(x,y)$-plane,

b) invariance under $S^z\to-S^z$,

c) local homogeneity of interactions, $h_{j,\,j+1}=h_{j+1,\,j}$.\\
\\
The most general anisotropic spin-one chain with nearest and next-nearest
neighbour interactions can then be written in the following form:\\
\begin{eqnarray}
{\cal H} & = & \sum_{j=1}^L h_{j,j+1,j+2} = \sum_{j=1}^L \frac{1}{2}(h_{j,j+1}
+
               h_{j+1,j+2}) + h_{j,j+2}, \label{Ham}\\
h_{j,j+1}& = &\alpha_0A^2_j + \alpha_1(A_jB_j + B_jA_j) + \alpha_2B^2_j
              +\alpha_3A_j + \alpha_4B_j(1+B_j)\\
         &   & +\alpha_5 \left( \left(S^z_j \right)^2 +
               \left(S^z_{j+1} \right)^2 \right), \nonumber\\
h_{j,j+2}& = &\tilde{\alpha_0}\tilde{A^2_j}
              + \tilde{\alpha_1}(\tilde{A_j}\tilde{B_j}+\tilde{B_j}\tilde{A_j})
              + \tilde{\alpha_2}\tilde{B^2_j}
              + \tilde{\alpha_3}\tilde{A_j}
+\tilde{\alpha_4}\tilde{B_j}(1+\tilde{B_j})\\
         &   & + \tilde{\alpha_5}\left( \left(S^z_j \right)^2 +
               \left(S^z_{j+2}\right)^2 \right) + c, \nonumber
\end{eqnarray}
with real parameters $ \alpha_j $, $ \tilde{\alpha_j} $ and a constant $c$.
We impose periodic boundary conditions. \
\\
The nearest (next-nearest) neighbour interactions $A_j$ and $B_j$
($\tilde{A_j}$ and $\tilde{B_j}$) are defined as\\
\begin {eqnarray}
A_j & = & S_j^{\perp}S_{j+1}^{\perp}
                = S_j^{+}S_{j+1}^{-}+S_j^{-}S_{j+1}^{+}
\hspace{1cm}\mbox{(transverse),}\\
B_j & = & S_j^{z}S_{j+1}^{z} \hspace{4.5cm}\mbox{(longitudinal),}
\end {eqnarray}
\begin {eqnarray}
\tilde{A_j} & = & S_j^{\perp}S_{j+2}^{\perp}
                  = S_j^{+}S_{j+2}^{-}+S_j^{-}S_{j+2}^{+}
\hspace{1cm}\mbox{(transverse),}\\
\tilde{B_j} & = & S_j^{z}S_{j+2}^{z} \hspace{4.5cm}\mbox{(longitudinal),}
\end {eqnarray}
with $S^{\pm} = (S^x \pm i S^y)/ \sqrt{2}$. Thus we have a model with
12 non-trivial parameters $ \alpha_j $, $ \tilde{\alpha_j}$, including a scale,
and an additional constant $c$. \\
\\
If all the $ \tilde{\alpha_j}$'s are equal to zero,
we obtain the nearest neighbour model investigated in \cite{zit93}.
For this 6-parameter model it was shown that the exact groundstate could be
found as an unique MPG in a 4-dimensional parameter sub-space.
For various reasons it is
interesting to know what will happen if next-nearest neighbour interactions
are considered. The question is for which interaction parameters the
MPG ansatz is still applicable and what
is the structure of the solution manifold.\\

\section{Matrix product ground states}
We shall be interested in the antiferromagnetic case of the model
where the ground state is characterized by $S_{total}^z=0$. We use the same
ansatz as
in \cite{zit93} to determine the ground state in the form of
a matrix product state.\\
\\
Denoting the $S_j^z$ eigenstates with eigenvalues 0 and $\pm$1 by $|0>_j$ and
$|\pm>_j$,
we define at each site $j$ a $2 \times 2$-matrix:
\begin{equation}\quad
g_j =   \left( \begin{array}{cc} |0>          \quad&  -\sqrt{a}|+> \\
                                 \sqrt{a}|->  \quad&  -\sigma|0> \end{array}
        \right)_{j}  \label{g1}
\end{equation}
with non-vanishing parameters $a$ and $\sigma$.
The latter will turn out to be $\pm1$.
For the global ground state of (\ref{Ham})
we use the ansatz
\begin{equation}
|\psi_0\left(a, \sigma\right)> = \mbox{Trace} \left( g_1 {\scriptstyle
\bigotimes}g_2
                                           {\scriptstyle \bigotimes} \ldots
                                           {\scriptstyle \bigotimes} g_L
\right), \label{mpg}
\end{equation}
where ${\scriptstyle \bigotimes}$ denotes a matrix multiplication of
$2 \times 2$-matrices with a tensor product of the matrix elements
\cite{zit93}.\\
By adjusting the constant $c$, we guarantee the condition
$h_{j,j+1,j+2} \geq 0$ (and thus ${\cal H} \geq 0$).
We demand the MPG (\ref{mpg}) to be a ground state of (\ref{Ham}) with
eigenvalue 0:
${\cal H}|\psi_0\left(a, \sigma\right)> = 0$.
This obviously is equivalent with
$ h_{j,j+1,j+2}|\psi_0\left(a, \sigma\right)> = 0$. As (\ref{mpg})
is a product state, it is sufficient to demand
\begin{equation}
\fbox{\parbox{7cm}{\[
h_{j,j+1,j+2} \left(
  g_j {\scriptstyle \bigotimes} g_{j+1} {\scriptstyle \bigotimes}g_{j+2}
                \right) = 0. \]}} \label{mpggl}
\end {equation}
Equation (\ref{mpggl}) means that the local interaction $ h_{j,j+1,j+2}$
acts upon all the four entries of the $g^{ \bigotimes 3}$ matrix.
It can be shown that (\ref{mpggl}) requires nine linear equations for the
parameters $\alpha_j$ and $\tilde{\alpha_j}$ which guarantee the
ground state property of $|\psi_0\left(a, \sigma\right)>$ under the condition
$h_{j,j+1,j+2} \geq 0$.
As we have $12 + 2 = 14$ parameters, namely
the coupling parameters $\alpha_j$, $\tilde{\alpha_j}$
and the two parameters $a$ and $\sigma$ of the ansatz (\ref{mpg}),
this means that MPGs can be constructed in a five-dimensional parameter
submanifold.
The uniqueness of the ground state (\ref{mpg}) is achieved by a strictly
four-fold degenerate
lowest eigenvalue zero of the local interaction $ h_{j,j+1,j+2}$ \cite{zit92}.
This condition determines the geometrical structure of the
five dimensional solution manifold.

\section{Results}

Explicit calculations show that (\ref{mpggl}) is satisfied under the following
conditions
\begin{displaymath}
\begin{array}{l@{\qquad}l}
\mbox{(0)} & \sigma\; = \; \mbox{sign}\alpha_3, \\[3pt]
\mbox{(i)} &  a(\alpha_0 +2\tilde{\alpha_3}) \; = \; \alpha_3 -
\alpha_1,\\[3pt]
\mbox{(ii)} & \alpha_2 \; = \; \alpha_0 a^2 - 2|\alpha_3| +
4\tilde{\alpha_3}, \\[3pt]
\mbox{(iii)} & \alpha_5 \; = \; |\alpha_3| + \alpha_0(1 - a^2) -
a^2\tilde{\alpha_3}, \\[3pt]
\mbox{(iv)} &  \tilde{\alpha_0} \; =\; - \tilde{\alpha_3},\\[3pt]
\mbox{(v)} &  -a \sigma \tilde{\alpha_0} \; =\; \tilde{\alpha_3} -
\tilde{\alpha_1},  \\[3pt]
\mbox{(vi)} &  \tilde{\alpha_2} \; = \; (2 - a^2)\tilde{\alpha_3}, \\[3pt]
\mbox{(vii)} &  \tilde{\alpha_5} \; = \; {\scriptstyle\frac{1}{2}}(a^2 -
4)\tilde{\alpha_3},\\[3pt]
\mbox{(viii)} &  \tilde{\alpha_4} \; =\;
{\scriptstyle\frac{1}{2}}(a^2-2)\tilde{\alpha_3}.
\end{array}
\end{displaymath}
As pointed out before, the conditions (i) to (viii) define eight linear
equations for the
interaction parameters $\alpha_j$ and $\tilde{\alpha_j}$ whereas (0)
specifies $\sigma$. The most striking effect is that
the ansatz (\ref{mpg}) imposes strict conditions upon the parameters for
the next-nearest neighbour interactions, equations (iv) to (viii): As soon as
$\tilde{\alpha_3}\neq0$,
all the other parameters $\tilde{\alpha_j}$ do not vanish either,
except for special values of the parameter $a$. In particular we find:
$\tilde{\alpha_0}\neq 0$. This  means that as far as the
next-nearest neighbour interactions are concerned biquadratic terms
cannot be neglected.\\
\\
As $\tilde{\alpha_3}$ approaches zero (limit of only nearest neighbour
interactions),
the equations (iv) to (viii) become irrelevant. Besides,
the $\tilde{\alpha_3}$-terms
in (i), (ii) and (iii) vanish.
Comparing the two situations, $\tilde{\alpha_3} =0$ and
$\tilde{\alpha_3} \neq 0$,
the following observation can be made:
Starting from a model
with 12 non-trivial parameters for nearest and next-nearest neighbour
interactions,
we remain with five free parameters, $\tilde{\alpha_3}$, $\alpha_0$,
$\alpha_3$, $\alpha_4$
and $a$, whereas for only nearest neighbour interactions ($\tilde{\alpha_3}
=0$)
we remain with four free parameters, $\alpha_0$, $\alpha_3$, $\alpha_4$
and $a$, on the basis of a model with 6 non-trivial interaction
parameters \cite{zit93}.\\
\\
The ground state property of (\ref{mpg}) can be guaranteed under the condition
$h  \geq 0$. The uniqueness of (\ref{mpg}) is valid if all the other
eigenvalues of the local interaction are strictly positive. The diagonalization
of $h_{j,j+1,j+2}$ yields a set of linear, quadratic and cubic inequalities for
the
parameters $\alpha_j$ and $\tilde{\alpha_j}$ (see the appendix and \cite{cl}).
Those inequalities define a five-dimensional parameter manifold with a complex
geometrical
structure. In the limit $\tilde{\alpha_j} \rightarrow 0$ this manifold turns
into
the four-dimensional sub-space defined by the conditions
$ \alpha_0 > 0$, $\alpha_4 > 0$, $\alpha_3 \neq 0$ and $a \neq 0$. A detailed
discussion
of this solution manifold of the nearest neighbour model can be found in
\cite{zit93}.\\
\\
For general $\tilde{\alpha_3} \neq 0$ there are  the following most significant
effects
(see also the appendix and \cite{cl}):\\
\\
We get a higher degree of freedom concerning the choice of the parameters
$\alpha_0$ and $\alpha_4$, in particular they can be set equal to zero
(\ref{ungl1}).
On the other hand the ansatz (\ref{mpg}) implies stricter conditions for
the parameters $\alpha_3$, $\alpha_1$ and thus $a$:
The parameter $|\alpha_3|$ must exceed a certain finite minimum value
(\ref{ungl2}) in contrast to
the simple condition:
$|\alpha_3| \neq 0$ for $\tilde{\alpha_3}=0$.
As in \cite{zit93} the parameter $a$ must satisfy the condition $a\neq 0$.
But in contrast to \cite{zit93}, depending on the choice of the values for
$\alpha_0$ and $\alpha_4$, $a$ must no longer be chosen arbitrarily
small or large (\ref{ungl2}, \ref{ungl3}).\\
\\
The ground state correlation functions can be calculated, using the transfer
matrix method
of \cite{zit92,zit93}. In the thermodynamic limit $ L \rightarrow \infty$  and
for
$r \geq2$ we obtain the same longitudinal and transverse 2-site correlation
functions
with exponential decay as in \cite{zit93}:
 \begin{equation} \fbox{\parbox{10cm}{\[
    <S_1^zS_r^z> = -\frac{a^2}{\left( 1 - |a| \right)^2}
                \left( \frac{1 -|a|}{1 + |a|} \right)^{r}
             = -\frac{a^2}{\left( 1 - |a| \right)^2} \quad \mbox{e}^{-
\frac{r}{\xi_l}}
             \] }}\label{sz1r}
\end{equation}
with the longitudinal correlation length:
\begin{equation}
    \frac{1}{\xi_l} = \ln \left| \frac{1 + |a|}{1 -|a|} \right|,
\label{longkor}
\end{equation}
\begin{equation}
   \fbox{\parbox{10cm}{\[
   <S_1^x S_r^x> \equiv <S_1^y S_r^y> = -|a| \left( \sigma + \mbox{sign}a
\right)
                        \left( \frac{-\sigma}{1+|a|} \right)^{r}
                        \] }} \label{sx1r}
\end{equation}
with the transverse correlation length
\begin{equation}
    \frac{1}{\xi_t} = \ln \left| 1+|a| \right|. \label{transkor}
\end{equation}
As the correlation functions depend on the overlap parameter $a$, the only
difference
between the cases $\tilde{\alpha_3}=0$ and $\tilde{\alpha_3} \neq 0 $ is given
by
equation (i) which describes the functional dependence of $a$ on the
interaction parameters.\\
Note that for $a \rightarrow 0$ there is a critical transition (diverging
correlation
lengths (\ref{longkor}),(\ref{transkor})) into a phase
where all the spins lie in the xy-plane: $ |\psi_0> \rightarrow |0000 \ldots>$.
For $ a \rightarrow \infty$ the MPG approaches the N\'{e}el states
$|\pm\mp\pm\mp \ldots>$.\\
\\
There is a finite gap to the excitations in the thermodynamic limit. This can
be understood
using the same arguments as in \cite{zit93}.
The MPG ansatz generates a "Haldane scenario" in a natural way.
The global ground state of (\ref{Ham})
is composed of local ground states in a way which is independent of the
system's size.
For this reason the properties of the finite system, especially the energy gap,
persist
in the thermodynamic limit $ L \rightarrow \infty$.
\\[10pt]

\newpage

\renewcommand{\theequation}{A.\arabic{equation}}
\section*{Appendix}
\setcounter{equation}{0}

The diagonalization of $h_{j,j+1,j+2}$ yields the following conditions
for the uniqueness of the MPG (\ref{mpg}):\\
\begin{equation}
\begin{array}{l@{\quad > \quad}l}
    \alpha_4 + \mbox{min}(\tilde{\alpha_3},\: {\scriptstyle
\frac{1}{2}}|\alpha_3|)
    & 0, \\[7pt]
    \tilde{\alpha_3} + {\scriptstyle \frac{1}{4}}|\alpha_3|
    & 0, \\[7pt]
    \alpha_0 +  {\scriptstyle \frac{1}{2}}|\alpha_3|
    & 0, \\[7pt]
    |\alpha_3| & 0 \\[7pt]
    2a^2\tilde{\alpha_3} + 2\alpha_4 +3|\alpha_3|
    & 0,
\end{array} \label{ungl1}
\end{equation}
\begin{equation}
\begin{array}{l@{\quad > \quad}l}
\tilde{\alpha_3} \left( 2 \left( a^2 - 2 \right)
                 \left( 2\tilde{\alpha_3} + \alpha_4 \right)
                 +|\alpha_3| \left(a^2 + 2 \right) \right)  + 2\alpha_4
|\alpha_3|
                 & 0, \\[7pt]
                 2\alpha_0 \left( 1+a^2 \right) +2a^2 \tilde{\alpha_3} +
3|\alpha_3|
                 & 0, \\[7pt]
       |\alpha_3| \left(\tilde{\alpha_3} + \alpha_0 \right)
                   -2 \tilde{\alpha_3} \left( 2\tilde{\alpha_3} + \alpha_0
\right)
                 & 0.
\end{array} \label{ungl2}
\end{equation}
\\
Using the following abbreviations:
\begin{equation}
\begin{array}{l@{\quad = \quad}l}
d_1 &    2(\tilde{\alpha_3} + \alpha_4), \\[7pt]
d_2 &    \alpha_4 + {\scriptstyle \frac{1}{2}}|\alpha_3| +\tilde{\alpha_3},
\\[7pt]
d_3 &    \tilde{\alpha_3}(3-a^2) + {\scriptstyle \frac{1}{2}}\alpha_0
+\alpha_4,\\[7pt]
d_4 &    |\alpha_3| + \tilde{\alpha_3}(a^2 - 2), \\[7pt]
d_5 &    {\scriptstyle \frac{1}{2}}(|\alpha_3| + a^2\alpha_0) +
         \tilde{\alpha_3}(a^2 - 1), \\[7pt]
d_6 &    |\alpha_3| - \tilde{\alpha_3},\\[7pt]
d_7 &    \alpha_0 + 2\tilde{\alpha_3},\\[7pt]
d_8 &    {\scriptstyle \frac{1}{2}}(\alpha_0 + |\alpha_3|)
+\tilde{\alpha_3}\\[7pt]
d_9 &    |\alpha_3|,\\[7pt]
d_{10} & a^2(\alpha_0 + \tilde{\alpha_3}),
\end{array}
\end{equation}
\begin{equation}
\begin{array}{l@{\qquad}l@{\qquad}l}
             n_1 \:=\: {\scriptstyle \frac{1}{2}}\alpha_0, &
             n_2 \:=\: {\scriptstyle \frac{1}{2}}\alpha_3, &
             n_3 \:=\: {\scriptstyle \frac{1}{2}}(\alpha_3 -\alpha_1),\\[7pt]
             n_4 \:=\: -\tilde{\alpha_3}, &
             n_5 \:=\: \tilde{\alpha_3}, &
             n_6 \:=\: a\sigma\tilde{\alpha_3},
\end{array}
\end{equation}

\parskip0pt
we further obtain:
\begin{equation}
\begin{array}{l@{\quad}l@{\quad}l}
  d_3 + d_5 & > & 0,\\[7pt]
 (d_3 + n_5)(d_5 - n_5) - n_3^2 & > & 0, \\[7pt]
  d_3 + d_5 + d_7 + d_9 &>& 0, \\[7pt]
 (d_3 + d_5) (d_7 + d_9) + d_3d_5 + d_7d_9 &  &\\
  + n_5(d_3 -d_5 -n_5) - 2(n_1^2 + n_2^2 + n_6^2) - 3n_3^2 &>& 0,\\[7pt]
  d_3(d_5 + n_5)(d_7 + d_9) + d_7d_9 (d_3 + d_5) &  &\\
  + n_3^2(4n_1 + 2n_5) + 4n_2n_3n_6 &  &\\
  -2n_1^2(d_5 + d_9) -2n_2^2(d_3 + d_7) - n_3^2(2d_3 + d_7 + 3d_9) &  & \\
  -n_5(d_7 + d_9)(d_5 + n_5) -2n_6^2(d_5 + d_7 + n_5) + 2n_5(n_2^2-n_1^2)&>& 0.
\end{array} \label{ungl3}
\end{equation}

\end{document}